\documentclass{appolb}
\usepackage{epsfig}
\usepackage{amssymb,amsmath,array}
\usepackage{wasysym}

\newcommand{\whizard}{\texttt{WHIZARD}}
\newcommand{\GeV}{{\ensuremath\rm GeV}}
\newcommand{\TeV}{{\ensuremath\rm TeV}}
\newcommand{\chp}{\tilde{\chi}^+}
\newcommand{\chm}{\tilde{\chi}^-}

\newcommand{\ME}{\mathcal{M}}
\newcommand{\ab}{{\ensuremath\rm ab}}
\newcommand{\eqn}{equation}

\newcommand{\wt}{\widetilde}
\newcommand{\fb}{{\ensuremath\rm fb}}

\begin{document}
\title{(N)LO Simulation of Chargino Production and Decay%
\thanks{Presented by T. Robens at the Cracow Epiphany Conference on LHC Physics, 4-6.1.2008 }%
}
\author{T. Robens
\address{Institut f\"ur Theoretische Physik E, RWTH Aachen, D-52056 Aachen, Germany}
\and
J. Kalinowski, K. Rolbiecki
\address{Institute of Theoretical Physics, University of Warsaw, PL-00681, Poland}
\and
W. Kilian
\address{Fachbereich Physik, University of Siegen, D-57068 Siegen, Germany}
\and
J. Reuter
\address{Physikalisches Institut, University of Freiburg, D-79104 Freiburg, Germany}
}
\maketitle
\vspace{-8mm}
\begin{abstract}
\noindent
We consider NLO chargino production and decays
at the ILC. For this, we present an NLO extension of the Monte Carlo
Event Generator \whizard~ including the NLO production. For photonic corrections, we use both a fixed
order and a resummation approach. The latter method evades
the problem of negative event weights and automatically includes leading higher order
corrections. We present results for cross sections and event
generation for both methods. As a first step
towards a full NLO Monte Carlo, we consider a LO implementation of the
chargino production and subsequent leptonic decay and investigate
the precision of the sneutrino mass determination by means of lepton energy distributions in chargino decays. The SM and SUSY
backgrounds are included in our study using full matrix elements as
well as smearing effects from ISR and beamstrahlung. Without using energy distribution fits, 
the sneutrino mass can be determined with an error in the percent regime.
\end{abstract}
\vspace{-2mm}
PITHA 08/08, SFB/CPP-08-20
\section{Introduction}
In many GUT models, the masses of charginos tend to be near the
lower edge of the superpartner spectrum, and can be pair-produced at
a first-phase ILC with c.m.\ energy of $500\;\GeV$. The precise
measurement of their parameters (masses, mixings, and couplings) is
a key for uncovering the fundamental properties of the Minimal Supersymmetric Standard Model (MSSM)~\cite{Aguilar-Saavedra:2005pw}. Regarding the experimental
precision which is in the percent regime at the ILC, off-shell
kinematics for the signal process, and the reducible and irreducible
backgrounds~\cite{Hagiwara:2005wg} need to be included as well as
NLO corrections for chargino production. Here we present the
inclusion of the latter \cite{Kilian:2006cj, Robens:2006np}. For
decay modes, we focus on the leptonic decay with electron and muon
in the final state. If sneutrinos decay invisibly
into the LSP and a neutrino, this channel provides tools to determine sneutrino masses. Such decays, common in many scenarios within the MSSM, preclude threshold scans since sneutrinos cannot be reconstructed directly. The only possibility to access the sneutrino mass in
such a case is to select a cascade decay where the precise
determination of kinematic distributions gives access to the
sneutrino mass. Although this idea has already been
exploited~\cite{Freitas:2005et}, a thorough study of how precise the
mass determination for the sneutrino in the environment of an ILC
can be has as yet not been made. We study the pollution effects of
all reducible and irreducible SM and SUSY backgrounds on the
visibility of the signal as well as the precision of the sneutrino
and the chargino mass measurements.  We restrict ourselves to areas
in SUSY parameter space where charginos are within reach of a 500
GeV ILC.

\section{Chargino production at LO and NLO}
\subsection*{Fixed order approach}
The total fixed-order NLO cross section is given by
\begin{eqnarray}
\sigma_{\rm tot}(s,m_e^2) &=& \sigma_{\rm Born}(s) +
  \sigma_{v+s}(s,\Delta E_\gamma,m_e^2) +
  \sigma_{\rm 2\rightarrow 3}(s,\Delta E_\gamma,m_e^2),\nonumber
\end{eqnarray}
where $s$ is the c.m.\ energy, $m_{e}$ the electron mass, and
$\Delta\,E_{\gamma}$ the soft photon energy cut dividing the photon
phase space. The 'virtual' contribution $\sigma_{v}$ is the
interference of the one-loop corrections \cite{Fritzsche:2004nf}
with the Born term. The collinear and infrared singularities are
regulated by $m_e$ and the photon mass $\lambda$, respectively. The
dependence on $\lambda$ is eliminated by adding the soft real photon
contribution $\sigma_{\rm s} \,=\,f_{\rm soft}\,\sigma_{\rm
Born}(s)$ with a universal soft factor $f_{\rm soft}(\frac{\Delta
E_\gamma}{\lambda})$ \cite{Denner:1991kt}. We break the `hard'
contribution $\sigma_{\rm 2\rightarrow 3}(s,\Delta E_\gamma,m_e^2)$,
i.e.\ the real-radiation process
$e^-e^+\rightarrow\chm_i\chp_j\gamma$, into a collinear and a
non-collinear part, separated at a photon acollinearity angle
$\Delta\theta_\gamma$ relative to the incoming electron or positron.
The collinear part is approximated by convoluting the Born cross
section with a structure function $f_{\rm
h}(x;\Delta\theta_\gamma,\frac{m_e^2}{s})$ \cite{Bohm:1993qx}:
\begin{eqnarray}
\sigma_\text{h,c}({\textstyle \Delta E, \Delta \theta, s})&
=&\int_{\Delta
E,0}^{E_{max},\Delta\theta}\,dx_{i}\,d\Gamma_{2}\,f_{\rm
h}(x_{i})|\ME_\text{b}|^{2}(x_{i},s),
\end{eqnarray}
where $x_{i}$ denotes the momentum fraction of the respective
incoming beam after photon radiation and $d\Gamma_{2}$ the two
particle final state phase space.
The non-collinear part is generated explicitly using exact three
particle final state kinematics.

The total fixed order cross section
is implemented in the multi-purpose event generator
\whizard~ \cite{Moretti:2001zz,Kilian:2007gr} using
a `user-defined' structure
function and an effective matrix element
\begin{eqnarray}
|\ME_\text{eff}|^2&=& (1+f_\text{soft}(\Delta E_{\gamma},\,\lambda))
\,|\ME_{\text{Born}}|^{2}\,+\,2\,Re(\ME_{\text{Born}}\,
\ME_{\text{virt}}^{*}(\lambda))\nonumber
\end{eqnarray}
which contains the Born part, the soft-photon factor and the Born-1
loop interference term. In the soft-photon region this approach runs
into the problem of negative event weights~\cite{Kleiss:1989de}: for
some values of $\theta$, the $2\to 2$ part of the NLO-corrected
squared matrix element is positive definite by itself only if
$\Delta E_\gamma$ is sufficiently large. To obtain unweighted event
samples, an ad-hoc approach is to simply drop events with negative
weights before proceeding further.

\subsection*{Resummation approach}
Negative event weights can be avoided by resumming higher-order
initial radiation using an exponentiated structure function
$f_\text{ISR}$~\cite{Gribov:1972rt,Skrzypek:1990qs}. In order to
avoid double-counting in the combination of the ISR-resummed LO
result with the additional NLO contributions
\cite{Fritzsche:2004nf}, we have subtracted from the effective
squared matrix element the soft and virtual photonic contributions
that have already been accounted for in $\sigma_{s+v}$. This defines
\begin{\eqn}
|\ME^{\text{res}}_\text{eff}|^2 \, = \,
|\ME_\text{eff}|^{2}- 2f_\text{soft,ISR}  \,|\ME_\text{Born}|^2
\end{\eqn}
which is positive even for low $\Delta E_{\gamma}$ cuts for all
values of $\theta$. Convoluting this with the resummed ISR structure
function for each incoming beam, we obtain a modified $2\to 2$ part
of the total cross section which contains all NLO contributions and
in addition includes higher order soft and collinear photonic
corrections to the Born/one-loop interference. This differs from the
standard treatment in the literature (cf.~\cite{Fritzsche:2004nf}) where higher order photon contributions
are combined with the Born term only (``Born+'').

The complete result also contains the hard non-collinear $2\to 3$
part convoluted with the ISR structure function:
\begin{eqnarray}
\lefteqn{\sigma_{\text{res,+}}=\int^{\Delta (E,\theta)} \,dx_{i}\,d\Gamma_{2}\,f_{\text{ISR}}^{(e^{+})}(x_{1})f_{\text{ISR}}^{(e^{-})}(x_{2})  |\ME^{\text{res}}_{\text{eff}}|^{2}}\nonumber\\
&&+\int_{\Delta (E,\theta)} dx_{i}\,d\Gamma_{3}\,f_{\text{ISR}}^{(e^{+})}(x_{1})f_{\text{ISR}}^{(e^{-})}(x_{2})|\ME^{2\rightarrow\,3}|^{2}
\end{eqnarray}
The resummation approach eliminates the problem of negative weights
such that unweighting of generated events and realistic simulation
at NLO are now possible in all regions of phase-space.
\section{NLO Chargino Production: Results}
\subsection*{Total cross section and relative corrections}
\begin{figure}
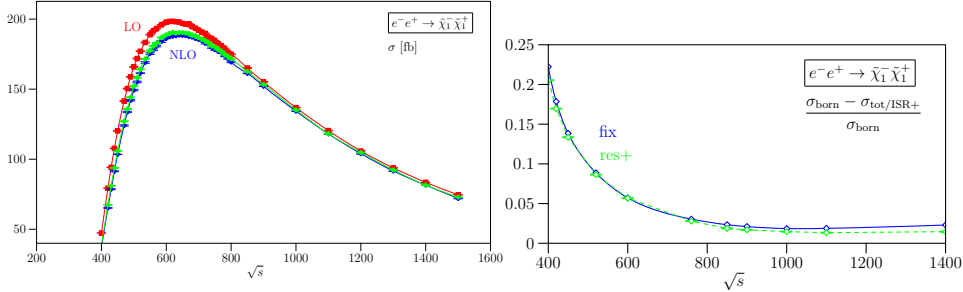

\centering
\includegraphics[width=.48\textwidth]{sigresum}
\hspace{5mm}
\includegraphics[width=.42\textwidth]{resumreldif} \vspace{5mm}
\caption{Total and relative cross section as a function of
$\sqrt{s}$. Left figure: Born (red, ``LO''), fixed order (blue,
``NLO'') and fully resummed (green, ``NLO'') total cross section,
right figure: relative  fixed order (blue, solid) and fully resummed
(green, dashed) higher corrections with respect to Born result}
\label{fig:sigtot}
\end{figure}
Figure \ref{fig:sigtot} shows the c.m.\ energy dependence of the
total LO and NLO cross section for chargino production for the
mSugra point SPS1a' \cite{Aguilar-Saavedra:2005pw} and the relative
corrections with respect to the Born result. The corrections are
mostly in the percent regime and can reach $20\%$ in the threshold
region.
\subsection*{Cutoff dependencies}
\begin{figure}
\centering
  \includegraphics[height=0.35\textwidth,width =0.6\textwidth]{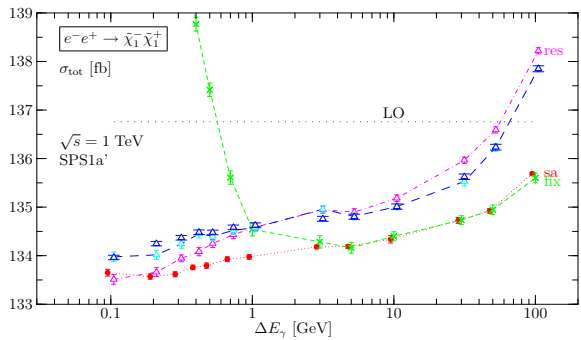}
  \caption{\label{fig:edep} {\small Total cross section dependence on $\Delta
    E_\gamma$: {\rm `sa'}
    (red, dotted) = fixed-order semianalytic result; {\rm `fix'} (green, dashed) = fixed-order
    Monte-Carlo result; {\rm `res'}
    (blue, long-dashed) = ISR-resummed Monte-Carlo result;
    (magenta, dash-dotted) = same but resummation applied only to
    the $2\to 2$ part. $\Delta\theta_\gamma=1^\circ$.
    LO: Born cross section.}}
\label{fig:edep}
\end{figure}
 Figure~\ref{fig:edep} compares the $\Delta E_{\gamma}$ dependence of
 the numerical results from
 a semianalytic fixed-order calculation with the Monte-Carlo
integration in the fixed-order and in the resummation schemes. The fixed-order Monte-Carlo result agrees with the semianalytic
result as long as the cutoff is greater
than a few $\GeV$ but departs from it for smaller cutoff values
because here, in some parts of phase
space, $|\ME_{\rm eff}|^{2}\,<\,0$ is set to zero. The semianalytic
fixed-order result is not exactly
cutoff-independent, but exhibits
a slight rise of the calculated cross section with increasing cutoff due to the breakdown of the soft photon approximation. For $\Delta E_\gamma=1\;\GeV$
($10\;\GeV$) the shift is about
$2\,{\permil}$ ($5\,{\permil}$) of the total cross section. The fully resummed result  shows an increase of about
$5\,{\permil}$ of the total cross section with respect to the
fixed-order result which stays roughly constant until $\Delta
E_\gamma>10\;\GeV$.  This is due to higher-order photon radiation.

For the dependence on the collinear cutoff $\Delta\theta_\gamma$,
the main higher-order effect is associated with photon emission
angles below $0.1^\circ$. For $\Delta
\theta_\gamma>10^\circ$, the collinear approximation breaks down.

\subsection*{Event distributions}
\begin{figure*}
\centering
  \includegraphics[height= 0.35\textwidth, width=.8\textwidth]{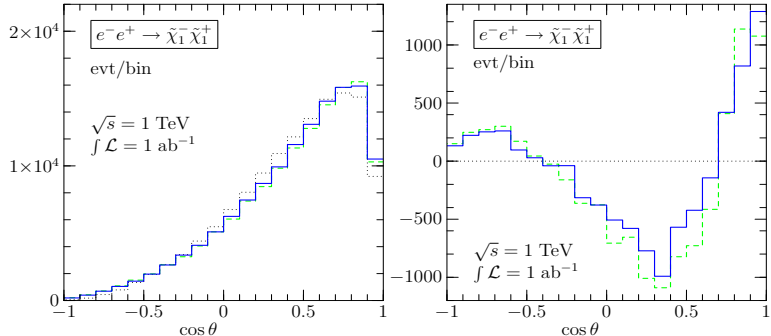}
  \caption{{\small Polar scattering angle distribution for an integrated
    luminosity of $1\;\ab^{-1}$ at $\sqrt{s}=1\;\TeV$. Left: total
    number of events per bin; right: difference w.r.t.\ the Born
    distribution.  LO (black, dotted) = Born cross section without
    ISR; fix (green, dashed) = fixed-order approach; res (blue, full)
    = resummation approach.}}
\label{fig:histth}
\end{figure*}

In Fig.~\ref{fig:histth} we show the binned distribution of the
chargino production angle obtained using a sample of unweighted
events. It demonstrates that NLO corrections to the angular
distribution are statistically significant and cannot be accounted
for by  a constant K factor.

\section{LO production and leptonic decays}
\subsection*{Signal and (MS)SM backgrounds}
In the second part of our work, we investigate the LO chargino
production and subsequent leptonic decay modes. To avoid large SM
backgrounds arising in the production of same flavor and opposite sign
lepton pairs, we consider
opposite flavor opposite sign lepton pairs in the final state:
\begin{\eqn}\label{eq:mainproc}
e^{+}\,e^{-}\,\longrightarrow\,\wt{\chi}^{+}_{1}\wt{\chi}^{-}_{1}
\,\longrightarrow\,\wt{\chi}^{0}_{1}\,\wt{\chi}^{0}_{1}\,e^{-}\,
\mu^{+}\,\nu_{\mu}\,\bar{\nu}_{e}.
\end{\eqn}
This channel is especially interesting  when the sneutrinos decay invisibly and the sneutrino pair-production channel is therefore experimentally inaccessible. In chargino pair-production and subsequent decays, however, the 
sneutrino and chargino masses can be determined from the edges of
the lepton energy distributions \cite{Freitas:2005et}. The
experimental signature of the signal (\ref{eq:mainproc}) is
\begin{\eqn*}
e^{+}\,e^{-}\,\longrightarrow\,e^{-}\,\mu^{+}\,+\,E_\text{miss}.
\end{\eqn*}
Therefore, all background processes where the missing energy results
from the emission of invisible particles ($\nu$'s,
$\wt{\chi}^{0}$'s) need to be considered. In addition, we have to
take into account processes where additional  particles are
emitted at very small angles and vanish in the beampipe. We
therefore investigate the SM and MSSM backgrounds  as listed in Table 1.
\begin{table}[!h]
\begin{\eqn*}
\begin{array}{l|l|l}
\text{\it ID}&\text{\it final state}&\text{\it most dominant process}\\
\hline\hline
\gamma\tau & e^{+}e^{-}e^{-}\mu^{+}\nu_{\mu}\,\bar{\nu}_{e}\nu_{\tau}\,\bar{\nu}_{\tau}&\gamma \text{-induced $\tau$ pair production (SM)}\\
WW & e^{-}\mu^{+}\nu_{\mu}\,\bar{\nu}_{e} & WW \text{ production (SM)}\\
\tau & e^{-}\mu^{+}\nu_{\mu}\,\bar{\nu}_{e}\nu_{\tau}\,\bar{\nu}_{\tau}&\text{$\tau$ pair production (SM)}\\
\tau W & e^{-}\mu^{+}\nu_{\mu}\,\bar{\nu}_{e}\nu_{\tau}\,\bar{\nu}_{\tau}\nu_{\tau}\,\bar{\nu}_{\tau}&\text{$\tau$  from $WW$ production (SM)}\\
\gamma W & e^{-}e^{+}e^{-}\mu^{+}\nu_{\mu}\,\bar{\nu}_{e}\nu_{\tau}\,\bar{\nu}_{\tau} &\gamma \text{-induced $WW$ production (SM)}\\
\tilde{\tau} &e^{-}\mu^{+}\nu_{\mu}\,\bar{\nu}_{e}\nu_{\tau}\,\bar{\nu}_{\tau}\wt{\chi}^{0}_{1}\wt{\chi}^{0}_{1}&\text{$\tilde{\tau}$ pair production (MSSM)}\\
\tilde{\tau}\nu_{\tau}
&e^{-}\mu^{+}\nu_{\mu}\,\bar{\nu}_{e}\nu_{\tau}\,\bar{\nu}_{\tau}\nu_{\tau}\,\bar{\nu}_{\tau}\wt{\chi}^{0}_{1}\wt{\chi}^{0}_{1}
&\text{$\tilde{\tau}\nu_{\tau}$ from $\wt{\chi}$ decays (MSSM)}
\end{array}
\end{\eqn*}
\caption{SM and MSSM background processes leading to $e^{-}\mu^{+}\,+\,E_\text{miss}$}
\end{table}

The values of the total cross sections for the SUSY parameter point
SPS1a' including beamstrahlung and initial state radiation before
the application of cuts\footnote{For numerical reasons, we always
include a collinear cut of $5^\circ$ for the outgoing $e^{-}$.} are
given in Table \ref{tab:xs}.
\begin{table}[!t]
\begin{\eqn*}
\begin{array}{l|l|l}
\text{\it ID}&\text{before cuts}&\text{after cuts}\\
\hline\hline
\text{signal}&3.940\,(8)&1.905\,(4)\\
\hline \hline
{\gamma\tau}&25495\,(4)&0.072\,(1)\\
{WW}&152.42\,(41)&0.794\,(2)\\
{\tau}&34.8\,(18)&0.024\,(1)\\
{\tau W}&2.978\,(9)&0.185\,(1)\\
{\gamma W}&2.192\,(12)&0.154\,(1)\\
\hline \hline
{\tilde{\tau}}&4.107\,(7)&1.146\,(2)\\
{\tilde{\tau}\,\nu_{\tau}}&2.74\,(9)&0.72\,(2)
\end{array}
\end{\eqn*}
\caption{\label{tab:xs} Signal and background total cross sections
before and after the application of background suppression cuts, for
SPS1a' and $\sqrt{s}\,=\,500\,\GeV$. ISR and beamstrahlung included.
All results are given in $\fb$, together with an integration
error.}
\end{table}
The most dominant background process is photon induced
$\tau^{+}\tau^{-}$ production; its cross section exceeds the
magnitude of the signal cross section by a factor $10^{4}$.
Similarly, SM background processes such as direct W and $\tau$
(pair) production are significantly larger than the signal, while
SUSY backgrounds are of similar size.

\section{LO production and decay: Results including cuts}
\begin{figure}
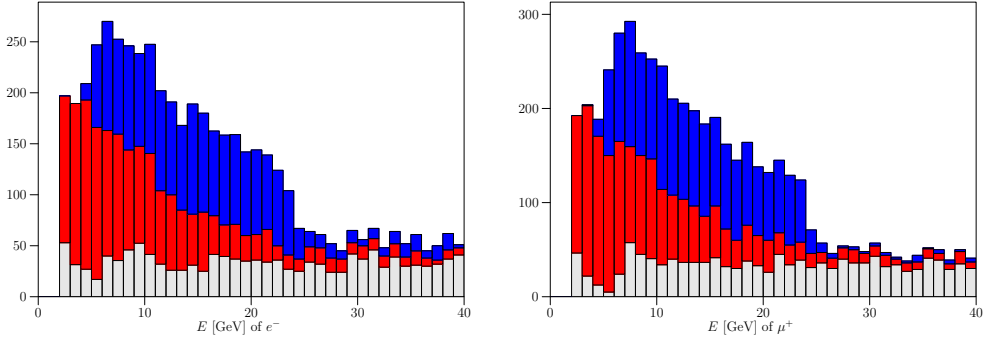

\centering
\includegraphics[width=.45\textwidth]{energiesetc.1}
\hspace{5mm} \quad
\includegraphics[width=.45\textwidth]{energiesetc.2} \vspace{5mm}
\caption{Energy distribution of electron (left) and muon (right) after cuts given in Table \ref{tab:cuts}. Grey: SM background; (photon-induced) $\tau^{+}\tau^{-}$ and $W^{+}W^{-}$ pair production. Red: SUSY background; $\tilde{\tau}\tilde{\tau}$ and $\tilde{\tau}\tilde{\tau}\nu_{\tau}\nu_{\tau}$ production. Blue: SUSY signal; $\wt{\chi}_{1}^{+}\wt{\chi}^{-}_{1}$ production and successive leptonic decays.}
\label{fig:eaftercuts}
\end{figure}

In order to suppress the SM and MSSM background, we apply the set of cuts given in Table \ref{tab:cuts}.
\begin{table}[!t]
\begin{eqnarray*}
p_{\perp}(e,\mu)\,\geq\, \textstyle 2\, \GeV, && \quad p_{\perp}(e)+p_{\perp}(\mu)\,\geq\,4\;\GeV,\nonumber\\
1\GeV\,\leq\,E(e,\mu)\,\leq\,40\GeV, && \quad -160^{o}\,\leq\,\Delta\,\phi\,\leq\,160^{o},\nonumber\\
15^{o}\,\leq\,\theta(e)\,\leq\,155^{o}, && \quad
25^{o}\,\leq\,\theta(\mu)\,\leq\,165^{o}
\end{eqnarray*}
\caption{\label{tab:cuts} Cuts applied for background suppression.
$\Delta \phi$ is the azimuthal separation angle of the lepton pair.}
\end{table}
The magnitudes for the total cross sections after the application of
these cuts are presented in Table~\ref{tab:xs}. The signal has been
reduced by roughly a factor $2$, while the dominant background
from photon induced tau pair-production is now suppressed by
$10^{6}$. In total, we obtain a signal/background ratio of $0.62$,
which, for an integrated luminosity of $1\,\ab^{-1}$, leads to a
$20\,\sigma$ discovery. Figure~\ref{fig:eaftercuts} shows the energy
distribution of the leptons after cuts have been applied. While the
SM processes lead to a flat background distribution which can be
easily subtracted, SUSY background processes are more challenging,
as they result in kinematic distributions similar to the signal.

\subsection*{Chargino and sneutrino mass determination}
In the case of quasi mass-degenerate sneutrinos and the lightest
charginos, the leptonic decay mode is dominated by intermediate on-shell sneutrinos:
\begin{\eqn}\label{eq:chdec}
\wt{\chi}^{\pm}\,\rightarrow\,l^{\pm}\,\tilde{\nu}_{l}\,\rightarrow\,l^{\pm}\,\nu_{l}\,\wt{\chi}^{0}.
\end{\eqn}
In these scenarios, sneutrino masses can only be determined from the
edges in the lepton energy distributions in chargino decays
(\ref{eq:chdec}). From on-shell relations, we obtain
\cite{Freitas:2005et}
\begin{\eqn}\label{eq:masses}
m_{\wt{\chi}^{\pm}}\,=\,\sqrt{s}\,\frac{\sqrt{E_\text{min}\,E_\text{max}}}{E_\text{min}+E_\text{max}},\quad
m_{\tilde{\nu}}\,=\,m_{\wt{\chi}^{\pm}}\,\sqrt{1-\frac{2\,(E_\text{min}+E_\text{max})}{\sqrt{s}}},
\end{\eqn}
where $E_\text{min/max}$ are the minimum/maximum lepton
energy. Taking naive readoff values for the energies, we have
(cf.~Fig.~\ref{fig:eaftercuts})
\begin{\eqn*}
E_\text{min}\,=\,4.5\,\pm\,1.0\,\GeV,\;E_\text{max}\,=\,24.5\,\pm\,2.0\,\GeV
\end{\eqn*}
and, using (\ref{eq:masses}),
\begin{eqnarray*}
&&m_{\wt{\chi}^{\pm}}\,=\,181\,\pm\,15\,\GeV\;(183.67),\;
m_{\tilde{\nu}}\,=\,170\,\pm\,14\,\GeV\;(173.52)\,,
\end{eqnarray*}
where the values in brackets are the nominal (input) values for
SPS1a'. Although the central values are in good agreement, the large
errors here call for a refined treatment. Alternatively, we take the
chargino mass from threshold scans; assuming
$\Delta_\text{thr}\,m_{\wt{\chi}}\,=\,1\,\GeV$, we then obtain
\begin{\eqn*}
m_{\tilde{\nu}}\,=\,172.7\,\pm\,1.3\,\GeV,
\end{\eqn*}
where the error is now in the percent regime. A fitting routine
for error reduction  down to a permille range using edge distributions including all
backgrounds is in the line of future work.

\section{Conclusions}
We have implemented NLO corrections into the event generator
\whizard\\for chargino pair-production at the ILC with several
approaches for the inclusion of photon radiation. A careful analysis
of the dependence on the cuts $\Delta\,E_{\gamma},\,\Delta\,\theta_\gamma$
reveals uncertainties related to higher-order radiation and
breakdown of the soft or collinear approximations. Careful choice of
the resummation method and cutoffs will be critical for a truly
precise analysis of real ILC data. The version of the program
resumming photons allows to get rid of negative event weights,
accounts for all yet known higher-order effects, allows for cutoffs
small enough that soft- and collinear-approximation artifacts are
negligible, and explicitly generates photons where they can be
resolved experimentally.

Additionally, we have investigated the
complete production and decay process at LO including several SM
and MSSM backgrounds. For this, we used a full matrix element
simulation including initial state radiation and beamstrahlung.
Efficient cuts reduce the dominant background by a factor  $10^{6}$
allowing a clear isolation of the signal. The leptonic decay
mode allows for mass determination of sneutrinos from the edges of
the lepton energy distributions. Taking the chargino mass from
threshold scans, the error of the sneutrino mass determination can
be reduced to the percent level.  The improvement of the mass determination precision using refined fitting routines and the combination of NLO production
and decay~\cite{Rolbiecki:2007cv,Fujimoto:2007bn} are in the line of
future work.

\section*{Acknowledgements}
This work was supported by the DFG SFB/TR9 "Computational Particle
Physics", the German Helmholtz Association, Grant VH-NG-005, the EU
Network MRTN-CT-2006-035505 ``Tools and Precision Calculations for
Physics Discoveries at Colliders", and Polish Ministry of Science
and Higher Education Grant No.~1~P03B~108~30.



\end{document}